\documentclass[onecolumn]{aa}
\usepackage{graphicx}
\usepackage{txfonts}

\newcommand {\hi} {H\,{\small I}}
\newcommand {\hii} {H\,{\small II}}
\newcommand {\kms} {km s$^{-1}$}

\newcommand {\mo}{$M_{\odot}$}
\newcommand{\ha}{H$\alpha$}
\newcommand{\nii}{[N~{\small II}]}
\newcommand{\sii}{[S~{\small II}]}

\begin{document}

\title{
Kinematics of the Ionised Gas in the Spiral Galaxy NGC 2403\footnote{
Based on observations made with the WHT operated on the
island of La Palma by the Isaac Newton Group in the Spanish Observatorio
del Roque de los Muchachos of the Instituto de Astrofisica de Canarias
}
}

\author{
        Filippo Fraternali\inst{1,2}, 
        Tom Oosterloo\inst{2}
        \and
        Renzo Sancisi\inst{3,4}
        }

\institute{
        $^1$ Theoretical Physics, 1 Keble Road, Oxford OX1 3NP, UK\\
        $^2$ ASTRON, P.O.\ Box 2, 7990 AA, Dwingeloo, NL\\
        $^3$ INAF, Osservatorio Astronomico, via Ranzani 1, 40127, Bologna, I\\
        $^4$ Kapteyn Astronomical Institute, University of Groningen, NL\\
        }

\date{Received ...; accepted ...}

\abstract{
We present a study of the kinematics of the ionised gas in the nearby
spiral galaxy NGC\,2403 using deep long-slit spectra obtained with the
4.2-m William Herschel Telescope. 
The data show the presence of a halo component of ionised gas that 
is rotating more slowly than the gas in the disk.  
The kinematics of this ionised halo gas is similar to that of the
neutral halo gas.  
On small scales, broad line profiles (up to 300 \kms\ wide) indicate
regions of fast outflows of ionised gas.  
We discuss these new results in the context of galactic fountain
models. 
\keywords{
galaxies: individual (NGC\,2403) -- 
galaxies: structure --
galaxies: kinematics and dynamics -- 
galaxies: halos -- 
ISM: kinematics and dynamics
} 
}

\authorrunning{Fraternali, Oosterloo, \& Sancisi}
\titlerunning{Kinematics of the Ionised Gas in NGC 2403}

\maketitle

\section{Introduction}

Recent studies of edge-on spiral galaxies have shown the presence of 
thick gaseous layers extending to large vertical 
distances from the plane of the disk.
Optical observations have revealed extraplanar emission from
ionised gas, frequently referred to as DIG (Diffuse Ionised Gas) 
(\cite{fer96}; \cite{hoo99}).
Such ionised layers are analogous to the Reynolds' layer of Warm Ionised Medium
(WIM) of our Galaxy (\cite{rey91}).
They can extend to several kpc above the plane (e.g.\ 15 kpc in
NGC 4631 (\cite{don95}); 5 kpc in NGC\,5775 (\cite{col00})),
but there are also galaxies that show little extraplanar emission
(\cite{ran92}). 
Deep \hi\ observations and a detailed modeling of the \hi\ layers
have revealed extraplanar gas also in the neutral
phase (NGC\,891 (\cite{swa97}); UGC\,7321 (\cite{mat03})).
Such extraplanar \hi\ is detected at several kpc from the plane,
up to 10$-$15 kpc, in NGC\,891 (\cite{fra03}).

The extraplanar gas has peculiar kinematics that has been
studied both for the neutral and the ionised component.
The modeling of the \hi\ layer in the edge-on galaxy NGC\,891 
(\cite{swa97}; \cite{fra03}), has shown that the extraplanar gas is
not co-rotating with the gas in the plane.
Instead, it rotates more slowly with a rotation velocity gradient in
the vertical direction of about 10$-$20 \kms\ per kpc and with larger
gradients in the inner regions of the galaxy.
Indications for such a rotation velocity gradient have
also been reported for the super-thin galaxy UGC\,7321 (\cite{mat03}). 
Long-slit spectroscopic observations of edge-on spiral galaxies
show similar results for the ionised component (NGC\,891
(\cite{ran97}); NGC\,5775 (\cite{ran00})).
The velocity gradient seems particularly remarkable in the edge-on
galaxy NGC\,5775 where the rotation velocity, measured in different
emission lines, seems to drop down to almost zero at a distance of
$\sim$~5$-$10 kpc from the plane.

The difference in rotation velocity between the planar 
and the extraplanar gas has been the key factor for revealing 
the latter also in galaxies
viewed at different inclination angles than edge-on.
Indeed, in such systems one can separate the ``lagging'' extraplanar
gas kinematically rather than spatially. 
NGC\,2403 is the first galaxy that has been studied for this purpose.
It is a nearby Sc spiral very similar to M\,33 and viewed at an
inclination angle of 60$^{\circ}$.
We have adopted a distance of 3.18 Mpc (\cite{mad91}).
NGC\,2403 is located in the far outskirts of the M\,81 group, at a projected distance 
of more than 1 Mpc from the centre of the group.
The closest object is the dwarf spheroidal DDO 44, at a projected
distance of 75 kpc, but without a clear physical connection with
NGC\,2403 (\cite{kar99}).
Recent \hi\ observations of NGC\,2403 have revealed the presence of tails 
in the \hi\ spectra at velocities lower than the rotation in the disk
(\cite{fra02a}).
Such tails (also referred to as "anomalous gas") have been successfully
reproduced by models with a thick layer (``halo'') of neutral
gas surrounding the thin disk while the thick layer has
a mean rotation velocity about 20$-$50 \kms\ lower than the thin disk 
(\cite{sch00}; \cite{fra01}).

The advantage of studying a galaxy viewed at intermediate inclination is that
one can investigate the presence of non-circular (inflow/outflow)
motions as well.
Indeed, in the case of NGC\,2403, in addition to the low rotation
velocity, also indications for a large-scale radial inflow (of
about 10$-$20 \kms) of the halo \hi\  have been found. 
This has been obtained by comparing the velocity fields of the
disk and of the extraplanar gas (\cite{fra01}).
Finally, the \hi\ data of
NGC\,2403 have also shown several gas complexes at quite
anomalous (non-circular) velocities.
These complexes have masses of about 10$^6-$10$^7$ \mo\ and 
velocities that depart from rotation by up to 150 \kms\ (300
\kms\ if the gas is moving vertically).
They form a population of high velocity clouds in the halo of
NGC\,2403
that are probably analogous to some of 
the HVCs of our Galaxy (\cite{wak97}).

The origin of the extraplanar gas in spiral galaxies is
still poorly understood.
One possibility is that 
it is produced by stellar activity in the disk and can be described by
the so-called galactic fountain model (\cite{sha76}; \cite{bre80}).
In such a model the gas is
thought to flow out of the disk in the halo region mostly as
ionised gas. 
Such gas can move to large heights above the plane 
(decreasing its rotation velocity) and to large distances
from its original radius conserving its angular momentum.
Then, after cooling and having become neutral, it is expected to fall
back onto the disk, eventually acquiring a radial inflow motion.
Therefore, in a galactic fountain process, one expects the early (outflow) 
stage to be mostly traced by ionised gas and the late (inflow) stage
by neutral gas.
As mentioned above, in NGC\,2403 there are indeed indications of a
large-scale radial inflow of neutral gas at a velocity of about 10$-$20 \kms\
which can be ascribed to the late stage of a galactic fountain.
However, this result, by itself, is not conclusive for the fountain
hypothesis.
Inflow is obviously expected also in the case of
accretion from the intergalactic medium (\cite{oor70}).
Therefore the study of the kinematics of ionised gas is crucial for
understanding the origin of the halo gas.

In this paper we present a study of the kinematics of the ionised gas 
in NGC\,2403 using long slit spectroscopy of strong emission
lines including \ha\ and \nii.
The main goal is to investigate the presence of an
ionised counterpart of the lagging halo of neutral gas (Section~3.2).
In particular, we search for outflows of ionised gas that could
match the observed inflow of neutral one in the hypothesis of a 
galactic fountain (Sections 3.3 and 4.1).
Finally, a comparison between the \hi\ and \ha\ 
rotation curves is also presented (Section~3.1).

\section{Observations and data reduction}

\begin{figure}
\centering
\includegraphics[width=66mm]{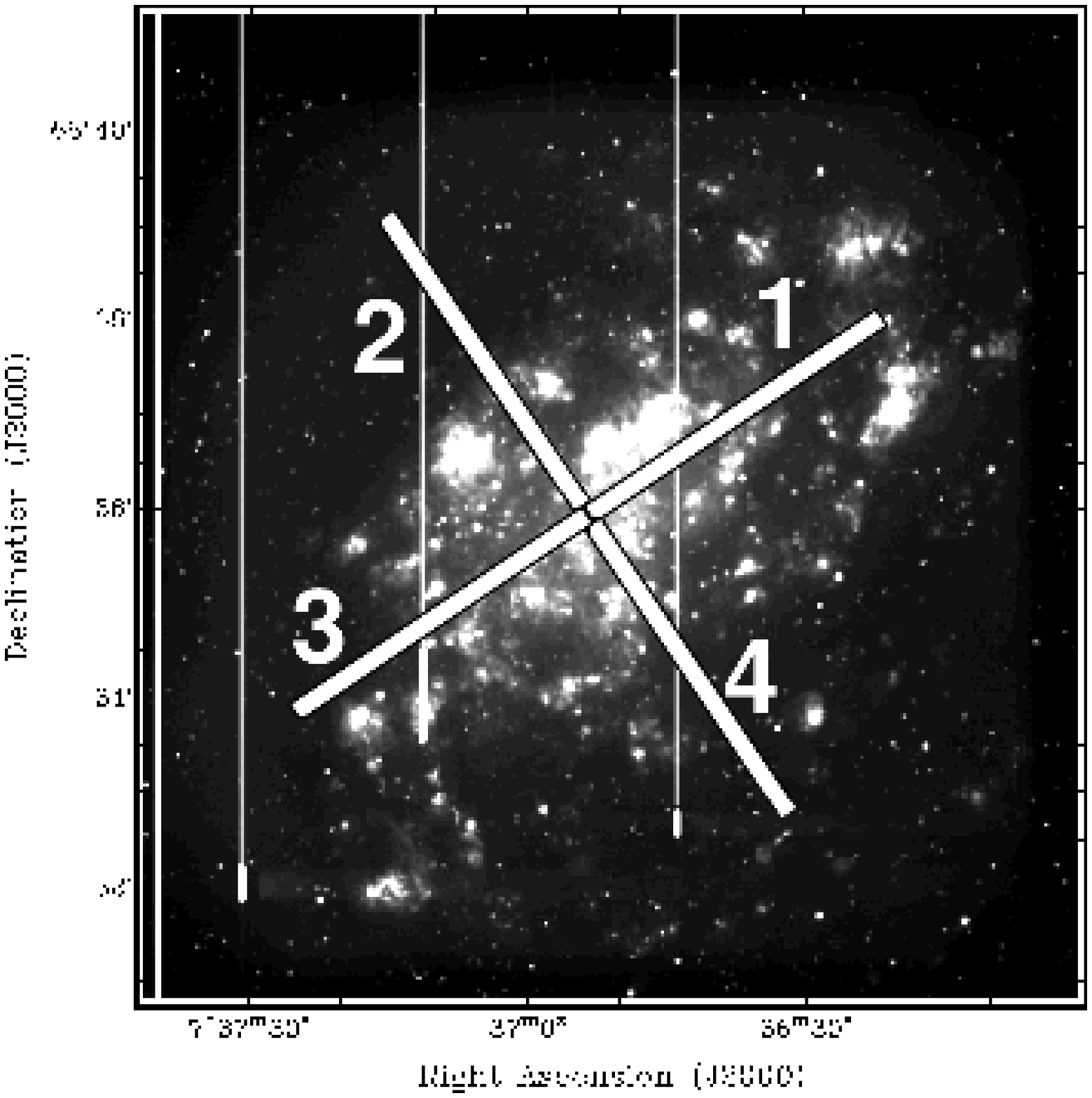}
\includegraphics[width=66mm]{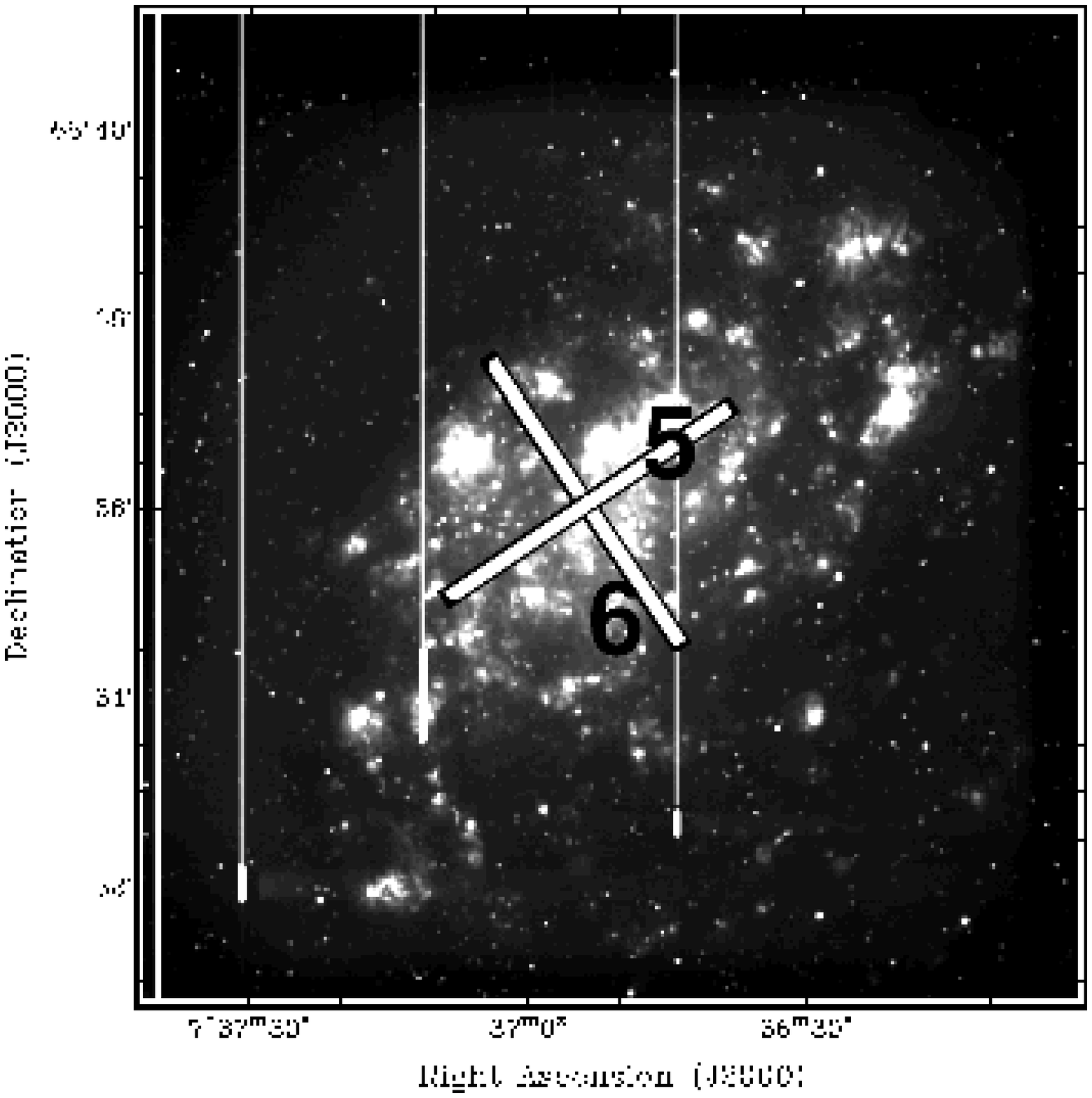}
\caption{
The 6 long-slits overlaid on an \ha\ image of NGC\,2403, 
taken from the archive of the
Canadian French Hawaiian (CFH) Telescope (\cite{dri99}).
The total exposure times in each position are given in Table~\ref{t_optobs}.
\label{f_slits}}
\end{figure}

Optical spectra of NGC\,2403 were obtained during three nights in January 2001
with the 4.2-m William Herschel Telescope at the Observatory Roque de los 
Muchachos in La Palma (Canary Islands) using 
the ISIS dual-beam spectrograph.
The weather conditions were generally good and the seeing was $\leq$1$''$ during the three nights.
The ISIS spectrograph on the WHT has two arms (``blue''
and ``red'') which can be exposed simultaneously.
The spectra of the blue arm range from 4500 to 5500 \AA{ }and include
H$\beta$ and [O{\small III}] (5007 \AA).
The spectra of the red arm (from 6200 to 6800 \AA) include
important emission lines: \ha\ (6563 \AA), \nii\ 
(6583 \AA, 6548 \AA) and the \sii\ doublet (6716 \AA, 6731 \AA).
The aim of our project is to study the presence of 
faint wings to the spectral lines;
thus, we mainly focus on the strong emission lines
\ha\ and \nii\ and use only red arm data.
The spatial scale of the red arm is 0$''$.32 pixel$^{-1}$ corresponding to 5
pc at the distance of NGC\,2403.
For the dispersion axis the scale is 0.4 \AA~pixel$^{-1}$ corresponding to
about 18 km~s$^{-1}$ pixel$^{-1}$ for lines in the red arm.
The velocity resolution is about 40 km~s$^{-1}$ which is four times lower than 
the velocity resolution of the \hi\ data (\cite{fra02a}).
We used a slit of 3.7 arcmin long $\times$ 1.2 arcsec wide
which we placed at different
positions along the major and minor axes of NGC\,2403
and obtained 48 2D-spectra (24 for each arm of the instrument).
Eight observations were taken along the major and the
minor axes of the galaxy with a total of six slit positions as shown in
Figure~\ref{f_slits}.
Each individual exposure was 45 minutes.
The central parts of the galaxy (R $<$ 1.8 $'$ $\simeq$~1.7 kpc) have 
obtained the longest integrations, ranging from 4.5 to 6.75 hours.
The other observational parameters are summarized in Table 1.

The standard data reduction (bias subtraction, flat-field division,
wavelength and flux calibrations)
was carried out with IRAF (Image Reduction and Analysis Facility) 
(\cite{val92}).
After wavelength calibration, small differences ($<$~2 \AA)
between the central wavelengths of sky lines in different exposures
were corrected by shifting all spectra in the dispersion axis to match
representative OH sky lines taken from the night sky atlas
(\cite{ost96}).
Then, the spectra taken at the same position and in the same night were 
combined to remove cosmic rays.
For the flux calibration Feige56,
LB1240, and Feige15 were used.
Finally, the continuum emission was subtracted from the combined
spectra by fitting a spline function of order 3 every row along the
dispersion axis.

The order 3 function was the best compromise between a smooth
continuum and low residuals.
 
\begin{figure}
\centering
\includegraphics[width=\textwidth]{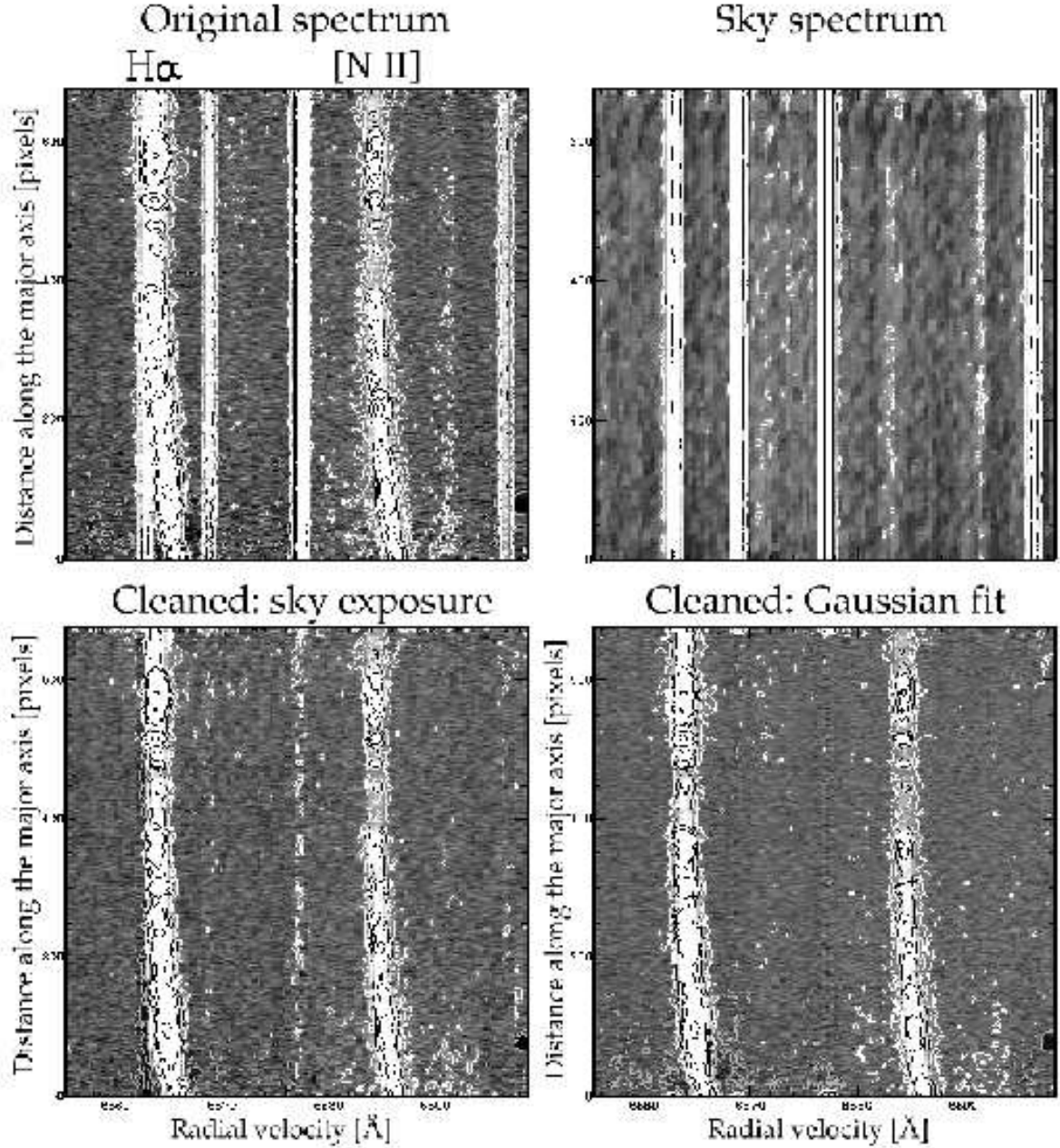}
\caption{Optical spectra for NGC\,2403 in the spectral range
6555$-$6600 \AA.
The upper left panel shows the original spectrum taken along the major
axis of NGC\,2403 in the N-W approaching side of the galaxy (slit~1 in
Figure~\ref{f_slits}), the centre of the galaxy is at the bottom.
The curvature of the emission lines (\ha\ and \nii) is caused by
the differential rotation.
The upper right panel shows a sky exposure in the same spectral range;
a spatial median filter was applied to this image.
The lower panels show the same spectrum as the upper left panel
after the sky subtraction.
Bottom left: sky subtraction obtained using the sky exposure.
Bottom right: sky subtraction with the Gaussian fit of the sky lines
as described in the text. 
The second method gives a better result.
The negative values in the bottom part of the spectra are probably 
caused both by stellar light absorption within NGC\,2403 and uncertainties
in the continuum subtraction.
The contours of all these spectra are 3, 6, 12 ... in units
of r.m.s\ noise of the original map, 1 $\sigma$ = 7.96 $\times$ 10$^{-19}$
 erg s$^{-1}$ cm$^{-2}$ \AA$^{-1}$ pixel$^{-1}$.
\label{f_skysub}}
\end{figure}

The critical step in reducing these spectroscopic data 
is the subtraction of the sky lines.
The presence of two bright sky lines close to
\ha\ and of the two faint lines embedding \nii\
(Figure~\ref{f_skysub}) 
requires a very careful sky subtraction.  Because
emission from NGC\,2403 is detected along the entire slit, the
standard technique of using the object-free regions of the CCD to make
an estimate of the sky emission cannot be used. 
Attempts to subtract the sky emission using separate exposures on
empty sky did not give satisfactory results (Figure~\ref{f_skysub},
bottom panel).
This is probably due to weak variations of the (absolute and relative)
strength of sky lines during the night.
To avoid these effects, we subtracted the sky lines directly from the
exposures of NGC\,2403.
Due to galactic rotation the galaxy emission shifts in velocity along
the slit (see \ha\ and \nii\ emission in Figure~\ref{f_skysub}). 
This allows us to find regions along the slit where it does not 
overlap with the sky lines.
We used these regions (typically 25 pixels) to extract the sky lines.
In the exposure shown in Figure~\ref{f_skysub}, for example, we used
the rows 1-25 for the line left to \ha\ emission and rows 301-325
for the one on the right.
We took the mean of these rows and fitted a Gaussian function to the 
line.

The intensity and the width of sky lines slightly vary along the
spatial axis.
In order to take this effect into account we have fitted Gaussian
functions to other sky lines (e.g.\ the bright line at about 6579 \AA)
in the same spatial regions (1$-$25 and 301$-$325).
Then we have repeated the fit along the whole spatial axis and used the 
resulting intensities and widths to scale the sky lines close to the
\ha\ emission.

A similar method was used to subtract the sky lines close to \nii\
but in this case, instead of fitting Gaussians,
we simply took the mean values along the spatial
regions free of \nii\ line emission. 
This method gives much better results than the direct
subtraction of the sky exposures (compare the bottom panels in 
Figure~\ref{f_skysub}).
We also successfully applied this method in the spectra along the minor
axis where the sky lines are isolated from the \ha\ and
\nii\ emission.

\section{Large-scale dynamics}

The main aim of this study is to investigate the presence of a
lagging halo of ionised gas in NGC\,2403.
This is motivated by our previous \hi\ studies which revealed
halos of neutral gas (about 1 tenth of the total \hi) rotating more
slowly than the disk.
The halo gas shows up in position-velocity (p-v) diagrams along the
major axis of the galaxy as a tail extending towards the systemic
velocity (\cite{fra01}).

\subsection{The optical rotation curve}

\begin{figure}
\centering
\includegraphics[width=66mm]{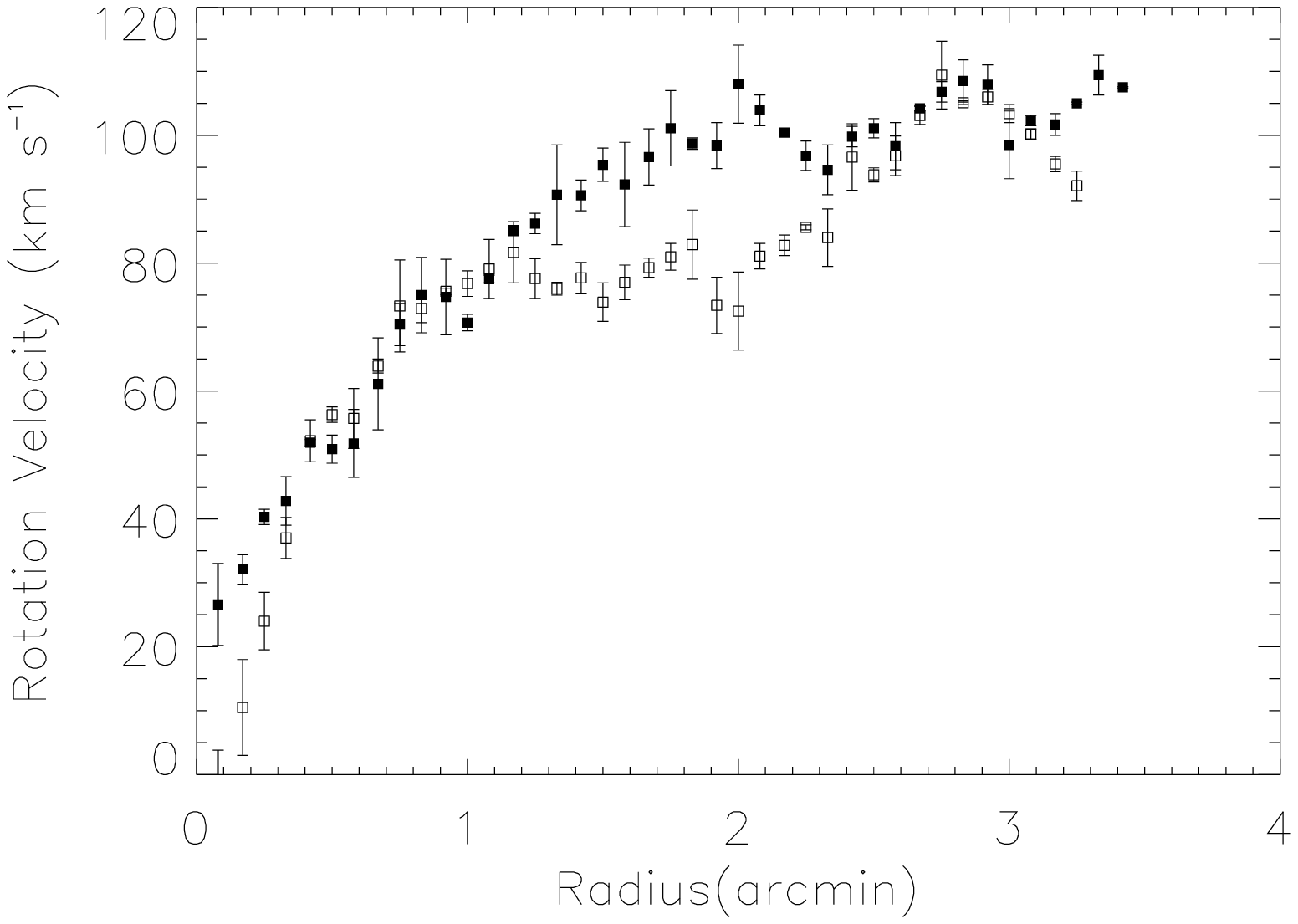}
\includegraphics[width=66mm]{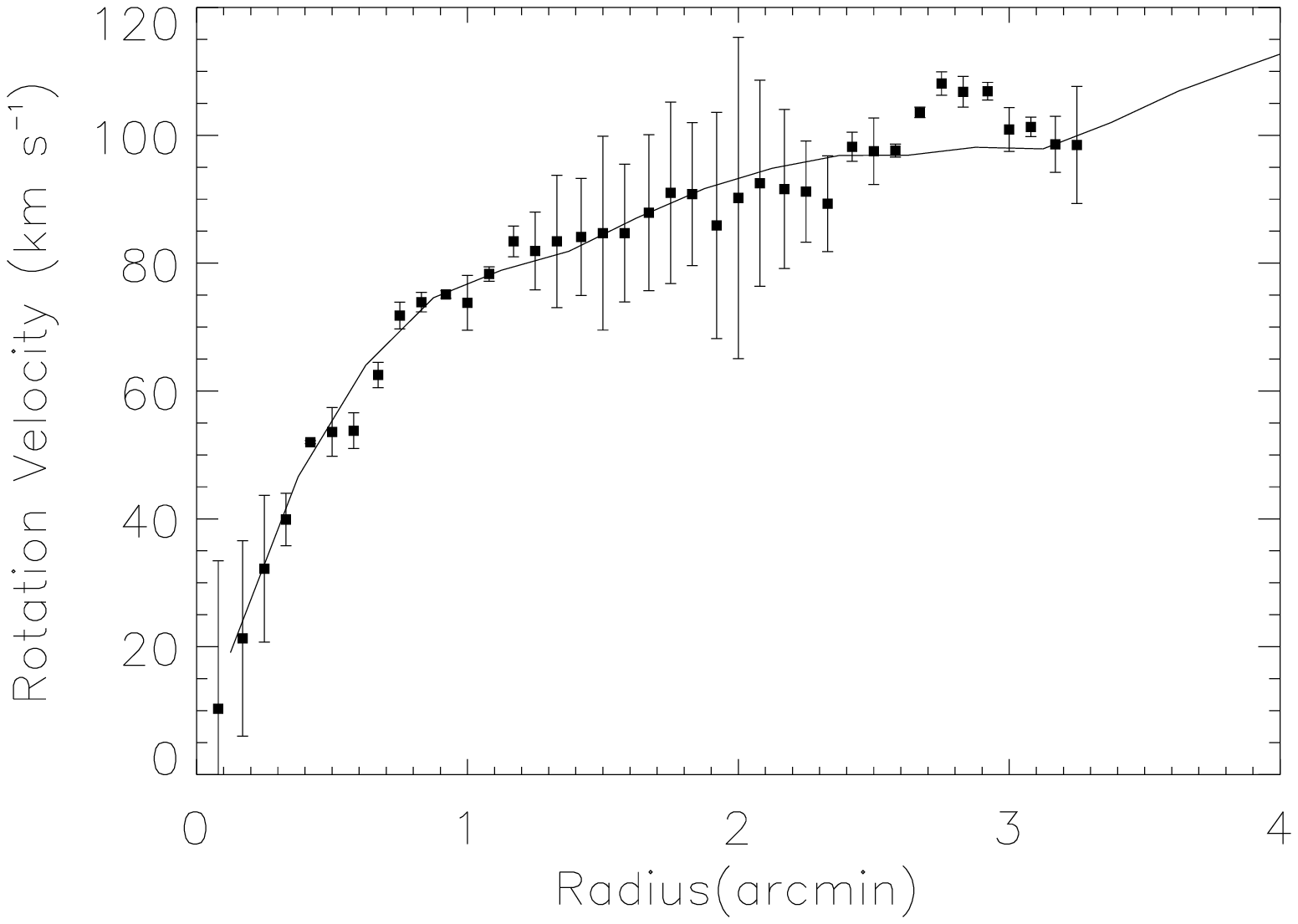}
\caption{
Optical \ha\ rotation curve for NGC\,2403.
The left panel shows the rotation velocity for the
receding (open squares) and approaching (filled squares)
sides of the galaxy. 
The right panel shows the final rotation curve.
In the right plot the error bars are obtained from differences between
the approaching and the receding sides and the continuous line shows
the \hi\ rotation curve. 
\label{f_rotcurs}}
\end{figure}

The optical rotation curves, in \ha\ and other emission lines,
are obtained from the Gaussian fits of the line profiles along the 
major axis described in Section 3.2.
Figure~\ref{f_rotcurs} shows the \ha\ rotation curve
(the \nii\ and \sii\ curves are very similar).
The left panel shows the rotation velocities for the receding and 
approaching sides,
the right panel shows the rotation curve for the whole galaxy.
In the right plot, the errors were estimated from the difference 
in velocity between the approaching and receding sides.
The continuous line shows the \hi\ rotation curve from \cite{fra02a}.
Clearly, the ionised and the neutral gas have the same kinematics.
The agreement between the \hi\ and \ha\ results is
remarkable if we consider that the two rotation curves
are obtained with different methods and 
from data with different resolutions.
There are systematic differences ($\leq$ 20 \kms) between the
approaching a receding rotation curves in the region 1$'-$2.5$'$, also
found in \hi.
These results from the \ha\ observations confirm the overall
regularity of the disk kinematics of NGC\,2403 already shown by the
\hi\ study.

\subsection{Lagging ionised gas}

\begin{figure}
\centering
\includegraphics[width=\textwidth]{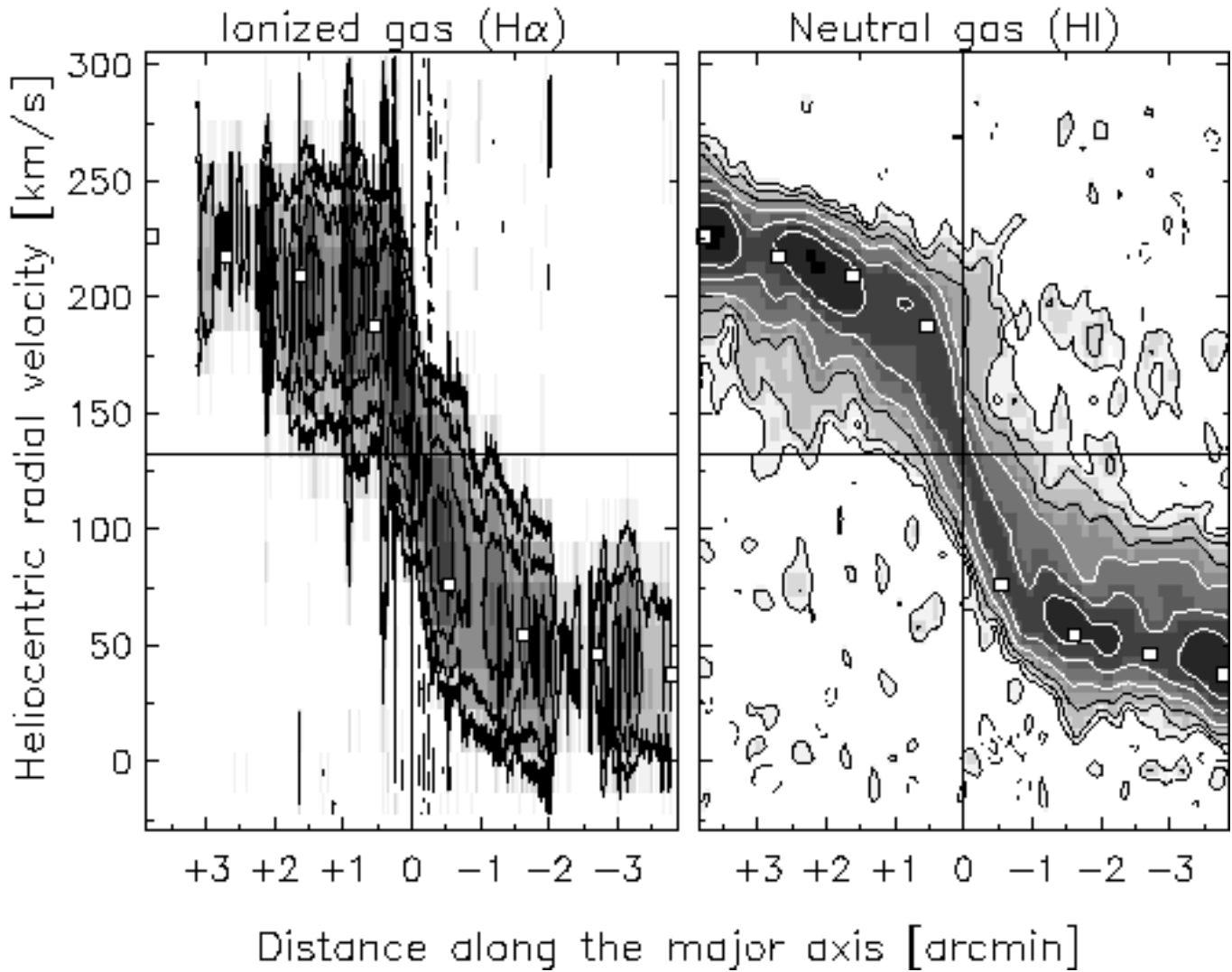}
\caption{
Position-velocity diagram along the major axis of NGC\,2403 for \ha\
(left) and \hi\ (right), both at full spatial and velocity resolution:
1$''$ $\times$ 40 km~s$^{-1}$ and 15$''$ $\times$ 10 km~s$^{-1}$
respectively. 
The \ha\ data are obtained by combining the observations along the major
axis (slit 1,3, and 5).
The \hi\ data are taken from \cite{fra02a}.
The white dots show the \hi\ rotation curve.
The contour levels, respectively for \ha\ and \hi\, are
$-$30, $-$9, $-$3, 3, 9, 30, 75, 225, 450 and $-$4, $-$2, 2, 4, 8,
16, 40, 80 in terms of r.m.s.\ noise.
\label{f_hhmax1}}
\end{figure}

Figure~\ref{f_hhmax1} shows the comparison between the \ha\ and \hi\ 
position-velocity diagrams (on the same scale) 
along the major axis of NGC\,2403.
Both are displayed at full spatial and velocity resolutions:
1$''$ $\times$ 40 \kms\ and 15$''$ $\times$ 10 \kms\ respectively for
\ha\ and \hi\ data.
The \hi\ data shown here are taken from \cite{fra02a}.
The \ha\ diagram was obtained by adding (without weighted averaging)
all the slit spectra taken along the major axis of the galaxy (see 
Table~\ref{t_optobs}).
The resulting exposure time along the major axis is larger (about a 
factor 2) in the central (R $<$ 1.8$'$) parts.
This causes the decrease of emission beyond R $\sim$ 1.8$'$.
The white dots are the \hi\ projected rotation curve showing a good
agreement between the overall kinematics of the 
neutral and ionised gas.
The central parts of this plot show negative values due to absorption
within NGC\,2403 and uncertainties in continuum subtraction.

\begin{figure}
\centering
\includegraphics[width=\textwidth]{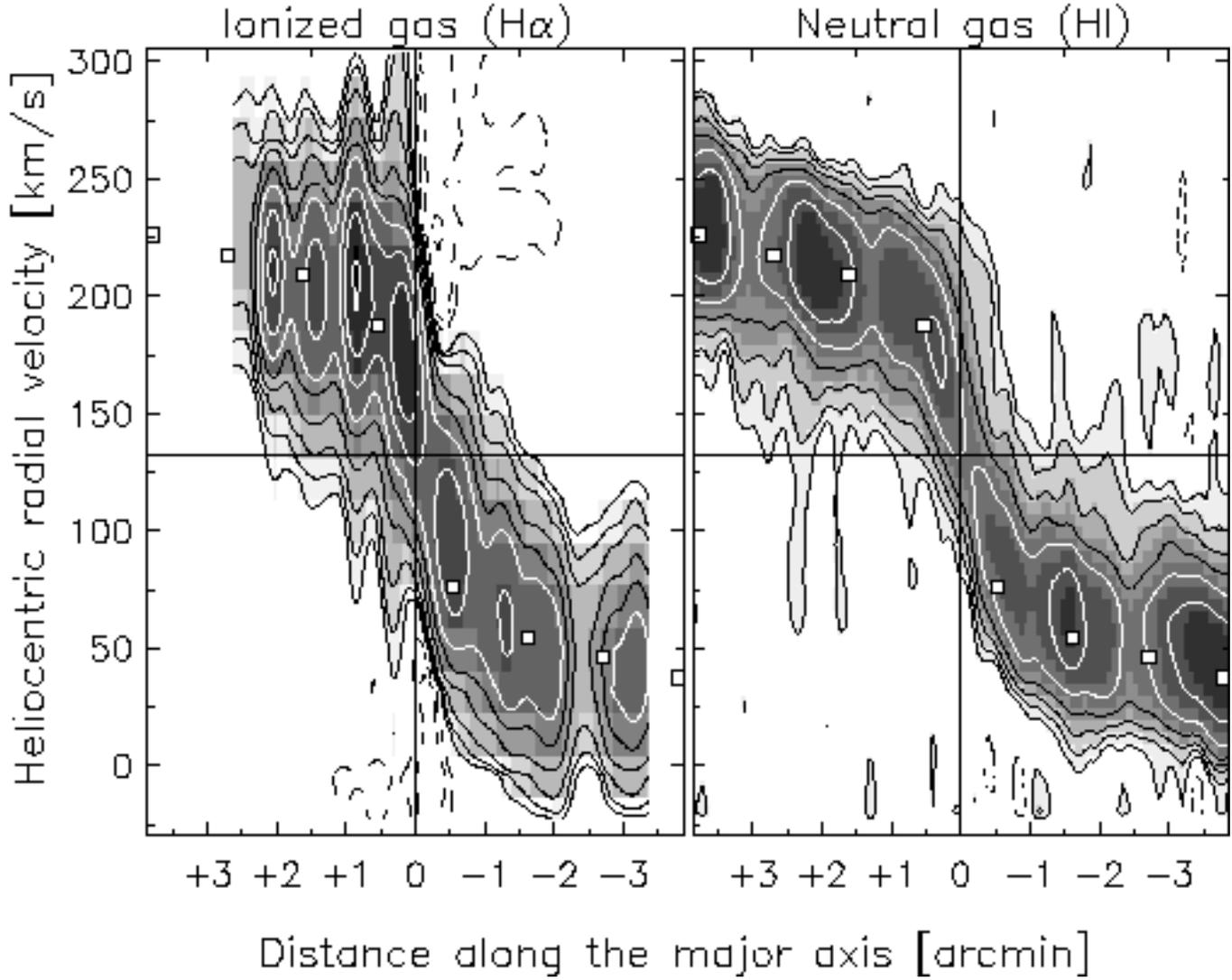}
\caption{
Position-velocity diagrams along the major axis of NGC\,2403 for
\ha\ (left panel) and \hi\ (right panel) at similar spatial and
velocity resolutions. 
The dots show the \hi\ rotation curve.
The contour levels, respectively for \ha\ and \hi\, are
$-$12, $-$6, $-$3, 3, 6, 12, 30, 66, 120, 300, 600, 1200 and $-$4, $-$2, 2, 4, 8,
16, 32, 64 in terms of r.m.s.\ noise of the smoothed data.
The negative levels in the central regions of the \ha\ plot are due
to continuum subtraction and stellar absorption in NGC\,2403.
\label{f_hhmax2}}
\end{figure}

The \hi\ p-v diagram in Figure~\ref{f_hhmax1} (right panel) 
shows tails of emission towards the systemic
velocity (horizontal line). 
As mentioned above, such tails are the signature of extraplanar gas 
rotating more slowly than the gas in the disk.
Is there a similar component also for the ionised gas?
At a first inspection of the \ha\ p-v diagram in Figure~\ref{f_hhmax1}
such a pattern is not clearly visible.
This may be partially due to the different resolutions of the ionised and
neutral gas data.
Figure~\ref{f_hhmax2} indeed shows the same p-v diagram as in
Figure~\ref{f_hhmax1} for \ha\ (left) and \hi\ (right) smoothed at
similar angular and velocity resolutions (15$''$ $\times$ 40 \kms). 
In this Figure the \hi\ tails are less visible than in
Figure~\ref{f_hhmax1} because of the lower velocity resolution.
Instead, the ionised gas (left panel) now shows some hints for 
these tails due to the higher signal-to-noise ratio obtained with the
spatial smoothing.

In order to further increase the signal-to-noise ratio, we
integrated all the line profiles in the direction along the major
axis.
Before doing this, we had to remove the overall rotation of the
galaxy.
To this purpose, we fitted Gaussian functions to each profile 
using the un-smoothed original data. 
The fitted intensities were used to normalize the profiles in flux
while the fitted velocities were used to shift them in velocity. 
This method was separately applied to the receding and the approaching 
sides of the galaxy for each emission line.

\begin{figure}
\centering
\includegraphics[width=\textwidth]{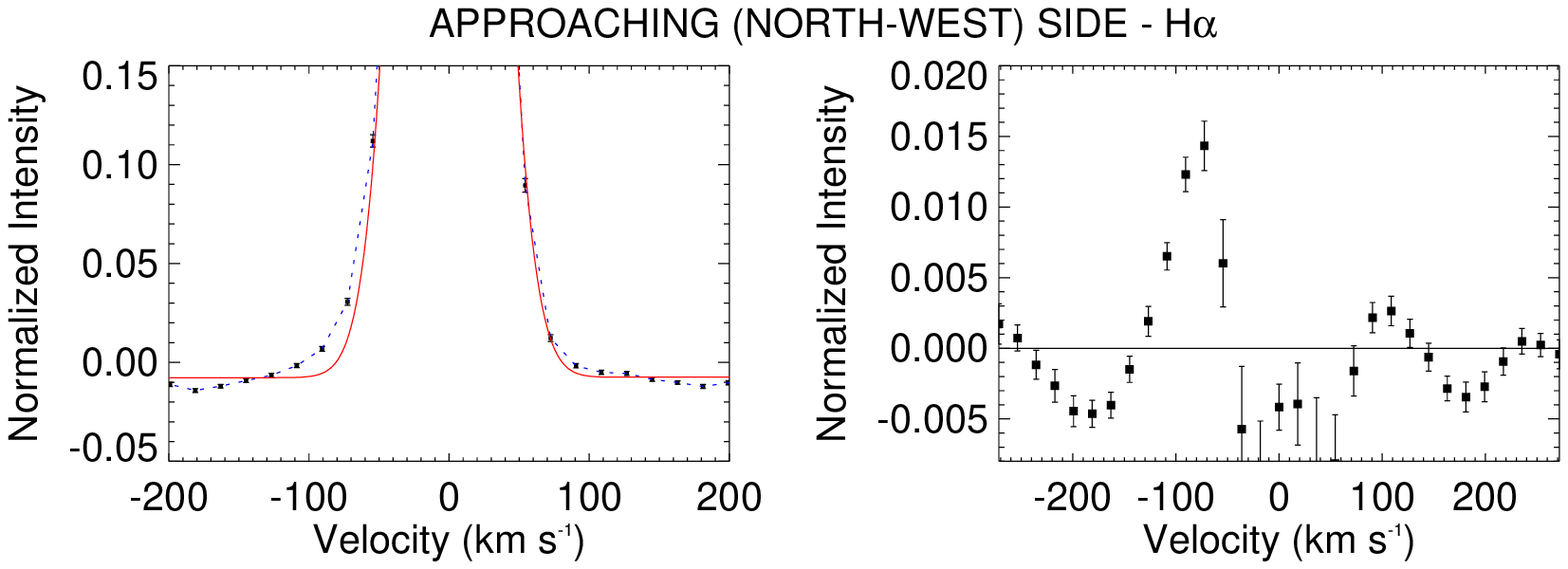}
\includegraphics[width=\textwidth]{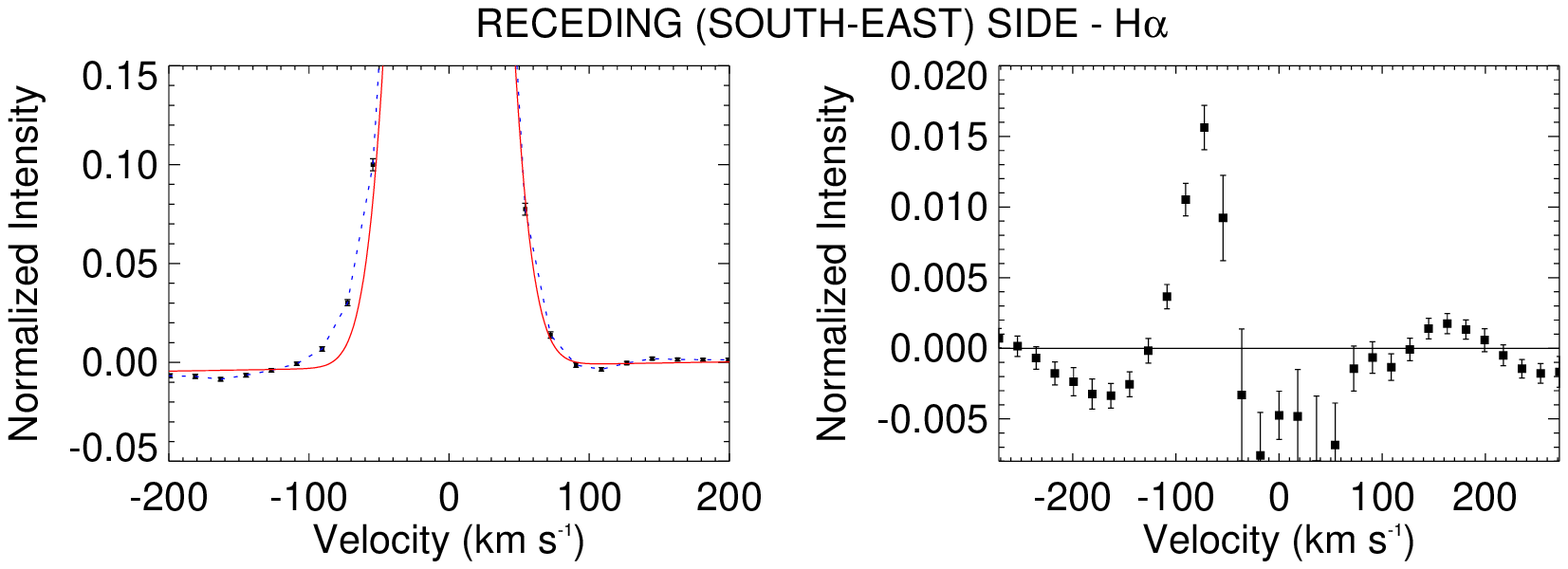}
\caption{
Composite \ha\ line profiles along the major axis of NGC\,2403
for both the approaching (upper panels) and receding (lower panels) sides
of NGC\,2403.
The continuous line indicates the Gaussian fit.
The tails are visible at low rotation velocities (negative velocities
in these plots).
Right panels show the residuals.
\label{f_coda_ha}}
\end{figure}

\begin{figure}
\centering
\includegraphics[width=\textwidth]{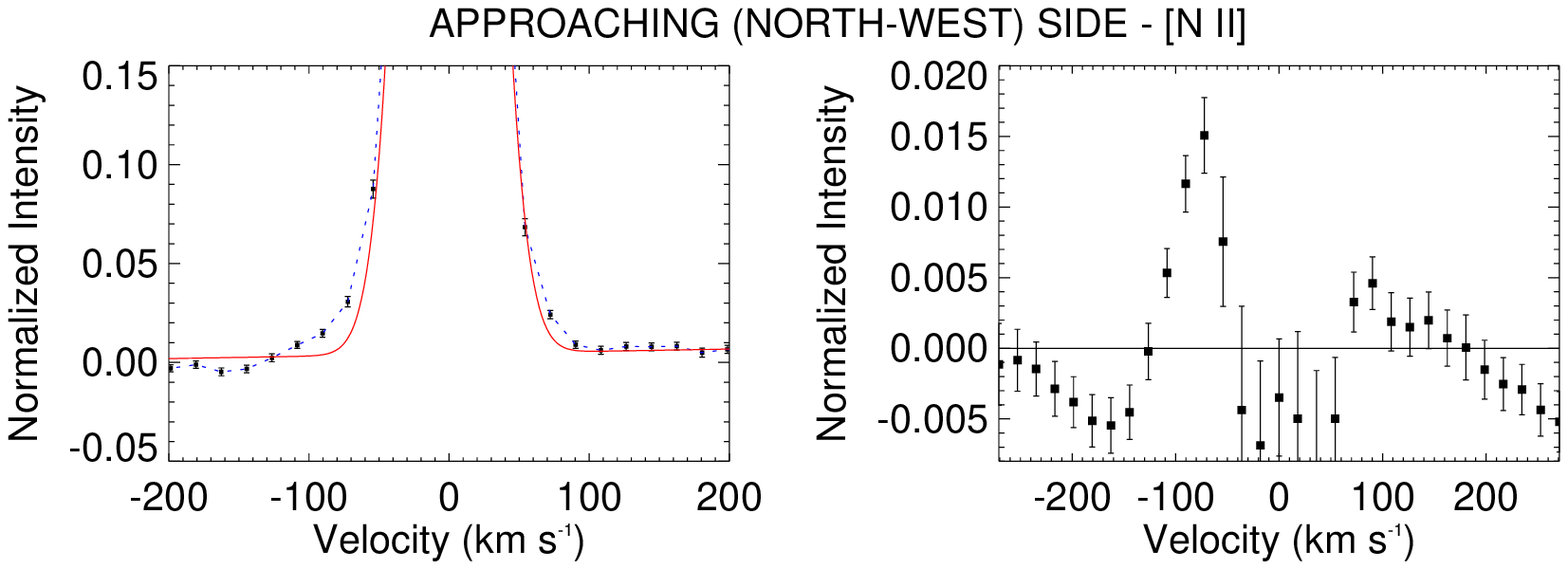}
\includegraphics[width=\textwidth]{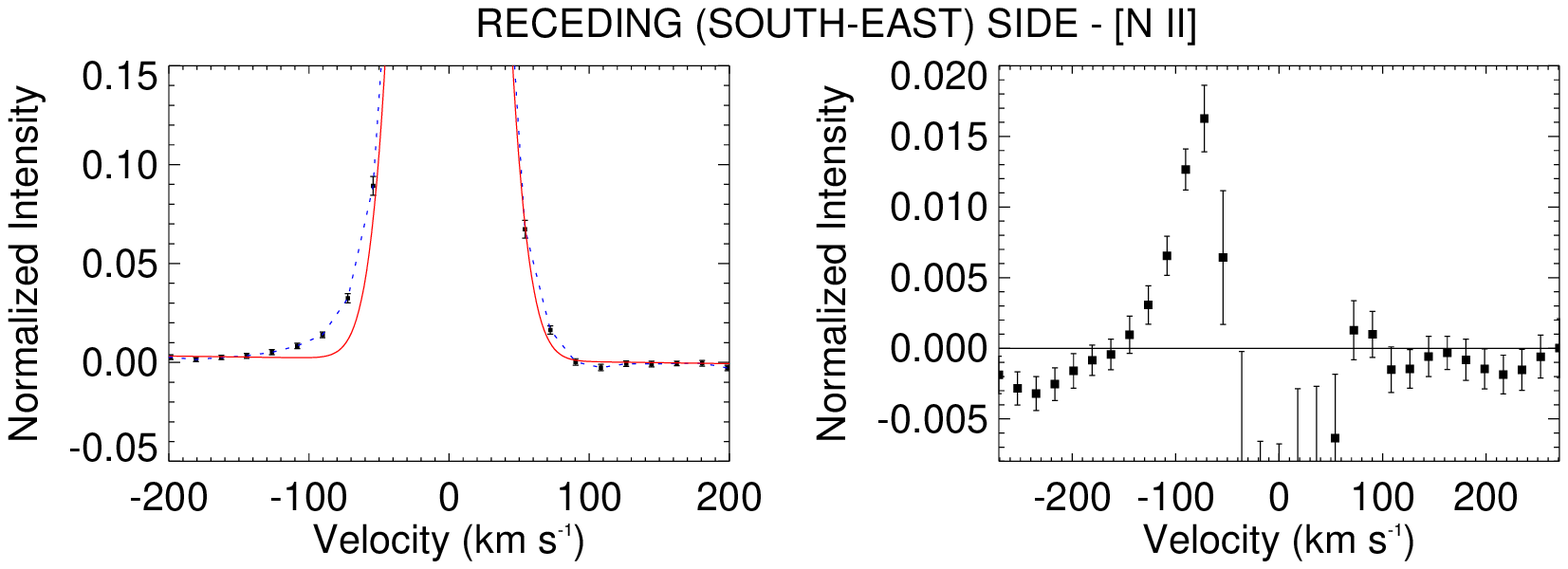}
\caption{
Composite \nii\ line profiles along the major axis of NGC\,2403
for both the approaching (upper panel) and receding (lower panel)
sides of NGC\,2403.
The continuous line indicates the Gaussian fit.
The tails are visible at lower rotation velocities (negative
velocities in these plots).
Right panels show the residuals.
\label{f_coda_nii}}
\end{figure}

The profiles resulting from this integration show that low rotation
velocity tails are present for the ionised gas.
The composite \ha\ line profiles for the receding and the
approaching sides of the galaxy are shown in Figure~\ref{f_coda_ha}
overlaid by a Gaussian fit.
The scale is compressed to better show the tails of emission.
The panels on the right show the residuals.
The velocity in the x-axis is $rotation$ velocity in a co-rotating
frame (rotation of the disk set to zero).
Figure~\ref{f_coda_nii} shows the same for the \nii\ line
(faint tails are also detected in \sii\ lines, not shown
here).
From these figures it is clear that the optical line profiles
do have tails of emission at low rotation velocities on both sides of the
galaxy.

Are these tails the indication of a lagging halo of ionised gas similar
to the \hi\ halo?
The ionised tails appear weaker than those observed in \hi\ 
line profiles.
From the residuals displayed in Figures~\ref{f_coda_ha} and
\ref{f_coda_nii}  
the ionised tail has a brightness of less than 2~\% of the line peak, 
while the typical ratio in the \hi\ profiles is about 10~\%
(\cite{fra02a}). 

The stellar absorption line at \ha\ can affect the shape of \ha\
emission line.  Using the results obtained by \cite{hoo03} who have
estimated the effect of stellar absorption lines on the gas emission
lines, we estimate this to  affect the emission from the tails by less
than 10$-$20 \%. Moreover, the fact that we obtain the same results
from both the H$\alpha$ and the \nii\ line indicates that the effect
is small.

For a proper comparison between the lagging neutral and ionised gas,
one should not consider the ratio in emission but the mass ratio.
The \hi\ emission is linearly related to the mass leading to a ratio
between the mass of the cold \hi\ disk and the halo gas of about 10\%
(\cite{fra02a}; \cite{sch00}).
Instead, for the ionised gas such a relation is rather complex.
Considering two scale heights for the thin disk and the lagging gas
$h_1$ and $h_2$ respectively, a 2\% ratio in emission gives a 14 
$\frac{h_1}{h_2}$ \% ratio in masses (assuming the same filling factor
for the two components).
Finally, the mass of the lagging ionised gas is probably 
under-estimated also because of the poor velocity resolution of the
optical spectra that partly hides the lagging gas from detection.

Concerning the kinematics, the ionised and neutral components
seem to match quite closely.
The tails of ionised gas are located (as those of \hi) only at low
rotation velocities on both sides of the galaxy.
The mean rotation velocity of the ionised lagging gas (about 80 \kms\
lower than the rotation in the disk) is somewhat larger than that
found in \hi\ but this result is strongly affected by the poor
spectral resolution.  
The ionised tails seem to extend up to more than 100 km~s$^{-1}$,
almost down to the systemic velocity of NGC\,2403, and such an
extension is very similar to the extension of the tails of neutral gas
in the central regions of the galaxy.
In some positions, the tails of ionised gas extend far beyond the
systemic velocity of the galaxy (see Section~4.1) as
also found for some high velocity \hi\ complexes (Figure~\ref{f_hhmax1}).

\begin{figure}
\centering
\includegraphics[width=\textwidth]{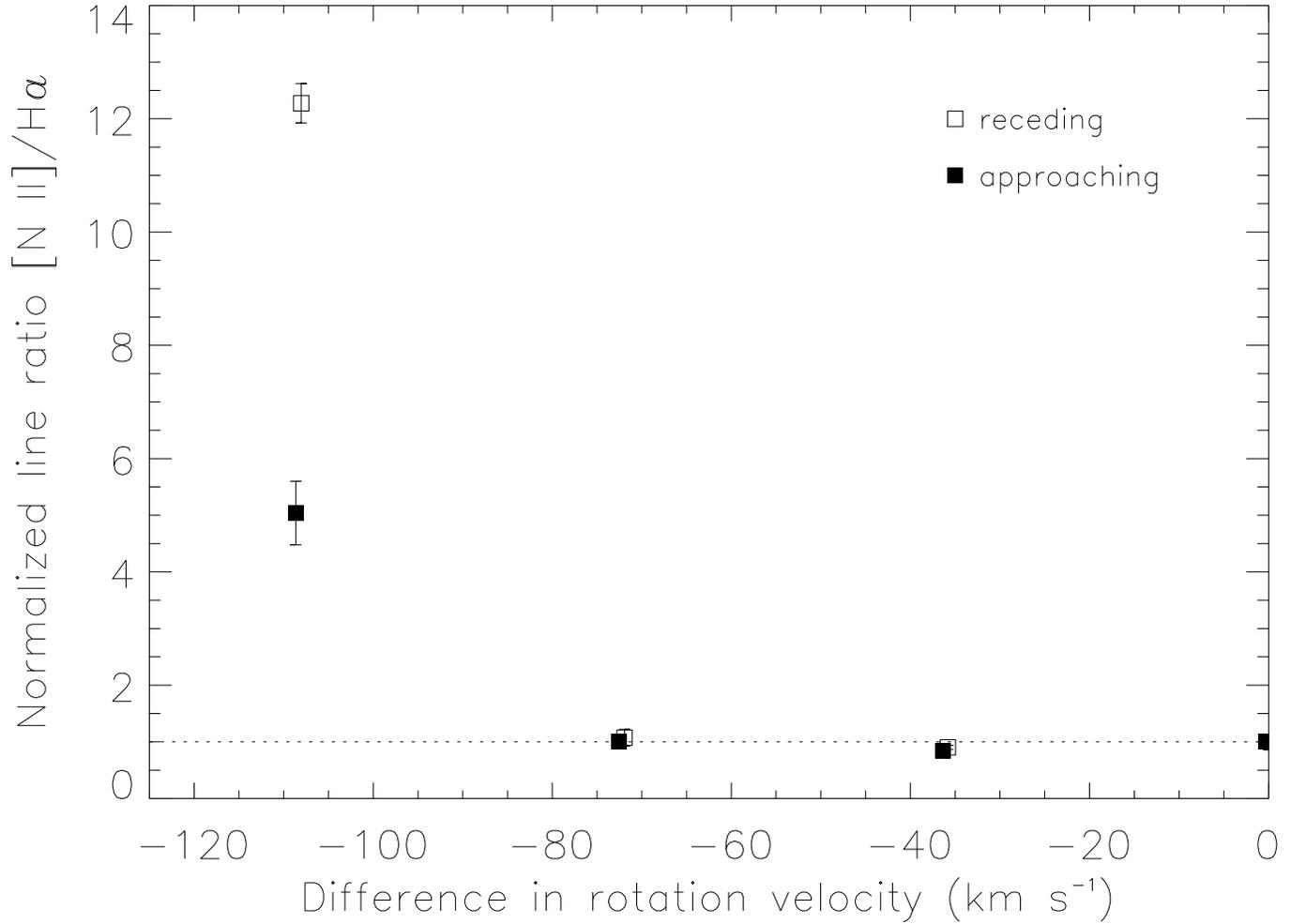}
\caption{
\nii/\ha\ intensity ratio versus velocity difference from the
rotation of the disk. The ratio is normalized at zero velocity. The
line ratio shows an increase at decreasing rotation velocities as
expected if this gas is located above the plane of the disk. 
\label{f_lineratio}}
\end{figure}

Finally, we have investigated the line ratios between different
emission lines.
Halo gas usually shows a higher \nii/\ha\ ratio with respect to gas
in the disk (e.g.\ \cite{hoo99}).
Figure~\ref{f_lineratio} shows a plot of this ratio versus rotation
velocity (rotation of the disk is set to zero).
This plot was obtained from the composite profiles shown in
Figures~\ref{f_coda_ha} and \ref{f_coda_nii}, therefore the ratio is
normalized at zero velocity.
The \nii/\ha\ ratio rises for gas at lower rotation velocity.
This clearly strengthens the interpretation that such a gas is located
above the plane and also suggests that the higher the
scale-height of the gas, the lower is its rotation velocity in
accordance with results obtained for edge-on galaxies (e.g.\ \cite{ran00}).

In conclusion, it appears that the tails at low
rotation velocity in optical emission lines found in NGC\,2403 represent a
lagging component of ionised gas which seems like the counterpart of the
lagging \hi\ halo. 
This is the first time that a lagging halo is detected in ionised gas
in a non edge-on galaxy.
The scale-height of such a component is unknown because of the
inclination of the galaxy.
However, an analogy with the extended and lagging DIG layers of edge-on
galaxies such as NGC\,5775 (\cite{ran00}) and NGC\,891 (\cite{ran97})
suggests scale-heights of a few kpc.

\subsection{Large-scale non-circular motions}

In Section~3.2 we have presented the evidence for a halo component of
ionised gas with lower rotation velocity with respect to the gas in
the disk. 
Here we start investigating the presence of non-circular motions. 
In this Section we concentrate on large-scale radial motions of the
ionised gas.
In Section 4.1 we discuss local features showing substantial vertical
motions.

Radial motions are revealed by p-v diagrams along and parallel to
the minor axis, where they produce an asymmetry between the near and 
the far side of the galaxy (\cite{fra01}).
Indeed, the minor axis is the $locus$ where there is no contribution by
the rotational velocity and where non-circular motions would show up.

\begin{figure}
\centering
\includegraphics[width=\textwidth]{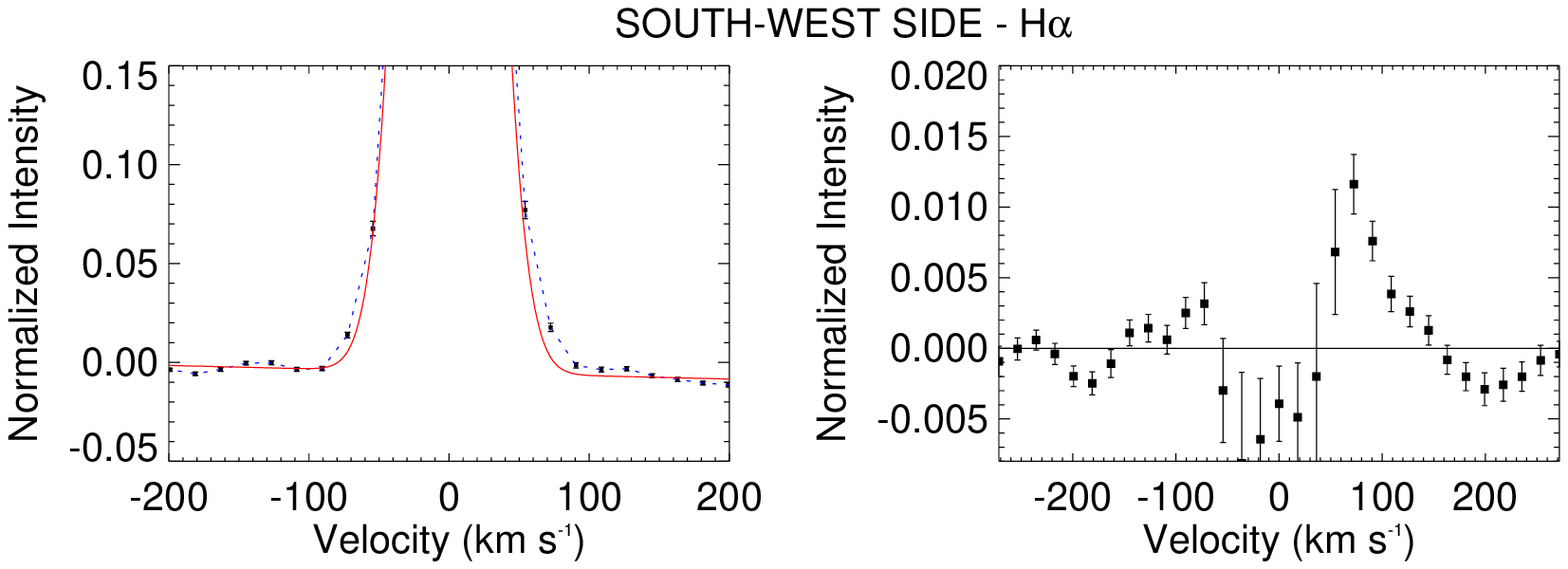}
\includegraphics[width=\textwidth]{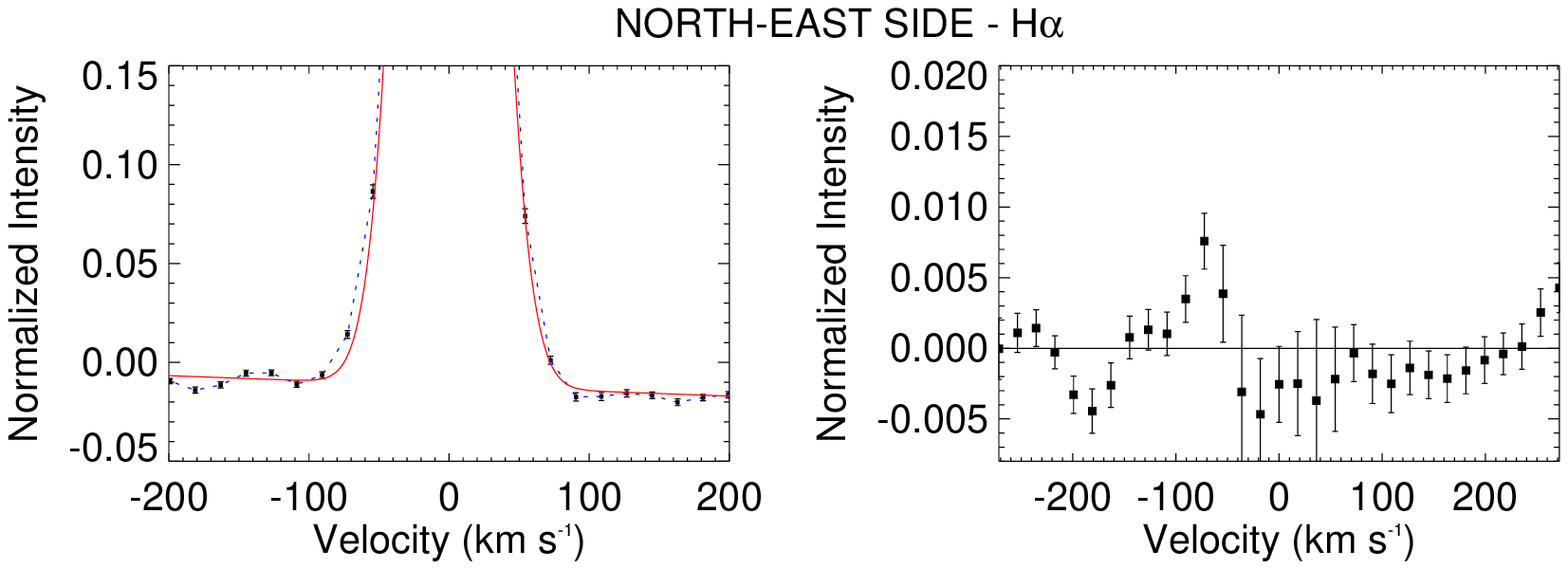}
\caption{
Composite line profiles for \ha\ along the minor axis of NGC\,2403 in
the S-W and N-E sides of the galaxy (left panels).
The continuous line indicates the Gaussian fit.
The right panels show the residuals.
Velocities are with respect to systemic.
The tail at high velocity in the S-W (near) side
and at low velocity in the N-E (far) side may indicate
a large-scale radial inflow.
\label{f_codamin_ha}}
\end{figure}

In order to look for such effects in the ionised component we have used
the slits along the minor axis of NGC\,2403 (slits 2, 4, and 6, see
Figure~\ref{f_slits}).
We have searched for tails of emission for the ionised gas 
using Gaussian fits as done for the major axis (Section 3.2).
Figures~\ref{f_codamin_ha} shows the composite \ha\ line profiles in both sides of the
galaxy.
The near side of the galaxy is the South-West side
for the spiral arms to be trailing and also considering dust absorption
features on the optical image.
Faint tails of emission are detected in
particular in the South-West (near) side of the galaxy,
at velocities higher than systemic.
In the North-East (far) side the tail is much fainter. 
However, this side has the lowest total exposure time (1.5
hrs in the outer and 4.5 in the inner part, see
Table~\ref{t_optobs}).

If one interprets these two \ha\ tails in terms of large-scale
non-circular motion, the ionised gas (as the \hi) would 
have a radial inflow towards the galactic centre.
Unfortunately, these tails are very faint (at least two times fainter
that those along the major axis) and a firm conclusion from this
analysis is not possible. 
No additional information comes from the study of \nii\ and \sii\ 
lines.

The kinematic position angle of NGC\,2403 is known with a precision of
1-2 degrees from \hi\ observations and we exclude that such tails
could have been produced by a misalignment between the slits and the
minor axis. 
This has also been tested by building models of galactic disks with 
different orientation of the minor axis.
The \hi\ velocity field of NGC\,2403 is regular and shows no sign of
wiggles between arm and inter-arm regions (\cite{fra01}). 
Moreover, the harmonic analysis of the \hi\ velocity field
(\cite{sch97}) shows that the disk of NGC\,2403 is axisymmetric to a 
fairly high degree.
Therefore we exclude that the tails of ionised gas at anomalous
velocity are produced by integrating through regions of the disk
with different kinematics.
The possibility that the gas in the tail and the bulk of the gas
spatially co-exists, with such a different kinematics, is quite
unlikely, these being collisional systems. 

If these tails are indeed real and part of a large-scale radial
inflow of ionised gas, this could be a fraction of the in-flowing
neutral gas which is being ionised by the stellar continuum in the
central parts of the galaxy. 
As for the neutral gas, the presence of inflow
motion may indicate that we are observing the last stage of the
fountain when the gas is falling back to the disk and moving inwards
(\cite{bre80}).
However, as already concluded for \hi\ (\cite{fra01}), the same
pattern would occur if the gas originates from accretion from the
intergalactic medium.
Therefore, the possible detection of large-scale radial inflow of
ionised gas does not confirm nor exclude the presence of a
fountain. 

On the other hand, if a galactic fountain is at work, we would 
expect to observe an outflow of ionised gas tracing the
initial stage of the process.
Here we have shown that such outflow is not detected as a large-scale 
radial motion.
However, on the small scales, there is evidence for significant 
outflow (vertical motions) of gas from the disk of NGC\,2403.
This is discussed in Section 4.1.

\section{Small-scale structures}

\subsection{Outflows of ionised gas}

Figure~\ref{f_maxc1_z1} (left panel) shows a spectrum along the major axis
(slit 5) of NGC\,2403 in the spectral range of \ha\ and
\nii.
At some locations, 
this spectrum shows a striking broadening of the profiles visible in each
line. 
They are labeled 2 and 3 and are at, respectively, about 0.5$'$ and
1$'$ from the centre of the galaxy (arrow).
The very broad feature detected only in \ha\ and labeled 1 is
discussed in Section 4.2. 
The bottom right panel shows the \ha\ line profile taken along
feature 2.
This profile shows long tails on both sides, at higher and lower
velocity with respect to the rotation (peak of the profile).
On the side of low velocities, the tail is much more pronounced 
and extends 200 km~s$^{-1}$ away from rotation.
On the high velocity side it reaches 300 km~s$^{-1}$ heliocentric
velocity. 

Both features 2 and 3 are spatially resolved.
Their physical sizes are about 50$-$100 pc.
In the \ha\ image (Figure~\ref{f_slits}) they correspond to compact
\hii\ regions of medium brightness.
There are no clear counterparts in \hi\ and
no X-ray counterparts are seen in the $Chandra$ image of NGC\,2403 
(\cite{fra02b}; \cite{sch03}).
The probable explanation of such local broadening of the line profiles is that
of strong vertical motions in regions of high star formation activity.
We searched for relations with known supernova remnants and
found a possible association (within our positional errors) with two 
SNRs for the feature 2 (numbers 14 and 15 in the catalogue of
\cite{mat97}) and no association for feature 3.

The presence of such features clearly points at outflows of
gas from the disk of NGC\,2403.
This can be part of the early stage of an ongoing galactic fountain.
The question is whether or not this process can be powerful enough to
account for all the extraplanar gas.
To answer it, we have estimated the flux density of
the high velocity gas in such features.
Consider first feature 2.
If we integrate only the emission at projected velocity differences
from rotation of more than 50 km~s$^{-1}$ we obtain an \ha\ flux
of  2 $\times$ 10$^{-15}$ erg s$^{-1}$ cm$^{-2}$ (luminosity of 2.4
$\times$  10$^{36}$ erg s$^{-1}$) and 
an emission measure ($EM$) of about 175 pc cm$^{-6}$, 
assuming a gas temperature of 10$^4$ K.
The emission measure $EM$ is related to the electron density $n_e$
$EM~(\mathrm{pc~cm^{-6}})=\int{n_e^2}{dl}$ (\cite{spi78}), this
leads to an $rms$ electron density for feature 2 
of $< n_e >_{rms}$ $=$ $1.6$ cm$^{-3}$, assuming a size of 70 pc for 
the region of emission.

The filling factor is defined as f $=$ $({\frac{< n_e
>_{rms}}{n_e}})^2$ where $n_e$ is the electron density of the ISM as
inferred using the ratio of emitted intensities of forbidden doublets.
We used the ratio of [S$_{\small II}$] lines and found a value of
$\frac{J6716}{J6731}$ $=$ 1.42 $\pm$ 0.02. 
Unfortunately this does not help us to clearly constrain n$_e$ and values
between 10 and 100 cm$^{-3}$ (for T=10$^4$ K) are all acceptable 
(\cite{ost74}).
We thus obtain a filling factor for the high velocity gas in the
feature 2 in the range f = 20$-$0.2 $\times$ 10$^{-3}$. 
Finally, assuming a spherical shape for the region of
emission, we get a value for the H$^+$ mass of
3.1 $\times$ $(\frac{f}{0.002})^{1/2}$ 10$^{2}$ \mo\ where $f$ is 
the filling factor. 
Such value for the mass correspond to a kinetic energy of about 4.5 $\times$
$(\frac{f}{0.002})^{1/2}$ $\times$ 10$^{49}$ erg s$^{-1}$.

For feature 3 we find an EM = 410 pc cm$^{-6}$, an $rms$ electron
density of $< n_e >_{rms}$ $=$ $2.3$ cm$^{-3}$ and 
a slightly higher filling factor f $\simeq$ 70$-$0.7
$\times$ 10$^{-3}$. 
The mass estimate in this case is 1.3 $\times$ $(\frac{f}{0.007})^{1/2}$ 
10$^{3}$ \mo\ and the kinetic energy 
4.7 $\times$ $(\frac{f}{0.007})^{1/2}$ $\times$ 10$^{49}$ erg s$^{-1}$. 

From our \hi\ observations, which indicate an \hi\ inflow velocity
of about 15 \kms, we estimate an infall rate 
$\dot{M}_{HI} \sim 0.9~\times~(\frac{L}{5~kpc})$ \mo\ yr$^{-1}$, where L
is the scale length of the process \cite{fra02a}.
Assuming a continuous cycle (i.e.\ assuming the same scale length)
between ionised and neutral gas, the contribution to the outflow rate
of ionised gas from features like the ones described above would be
$\dot{M}_{H+}$ $\approx 0.5-2.0$ $~\times~(\frac{L}{5~kpc})$ $\times$
10$^{-5}$ \mo\ yr$^{-1}$ with filling factors in the range $f$ =
1$-$10 $\times$ 10$^{-3}$.
Therefore one would need tens of thousands of such features to match
the observed inflow of \hi\ gas.
Note that this latter value does not depend on the chosen scale
length but only on the ratio between the mass and the velocity of the
out/in-flowing gas.
With the available data, it is hard to say whether this can be the case 
in NGC\,2403.
With our slits we have sampled a very small fraction of the disk of
NGC\,2403 ($\leq$ 1\%).
Moreover, none of our slits intercepts very bright \hii\ regions which
can easily be hundreds times more efficient than the two
detected here.
Finally, part of the outflowing gas can also be
hotter than the ionised gas observed in recombination lines
(\cite{fra02b}).

To sum up, we have detected local vertical outflows of ionised gas
that possibly feed a galactic fountain process in NGC\,2403.
Such a process could be responsible for the origin of the halo gas
if the galaxy hosts several thousands outflow regions
like the two reported here.

\subsection{Stellar outflow or Supernova?}

\begin{figure}
\centering
\includegraphics[width=\textwidth]{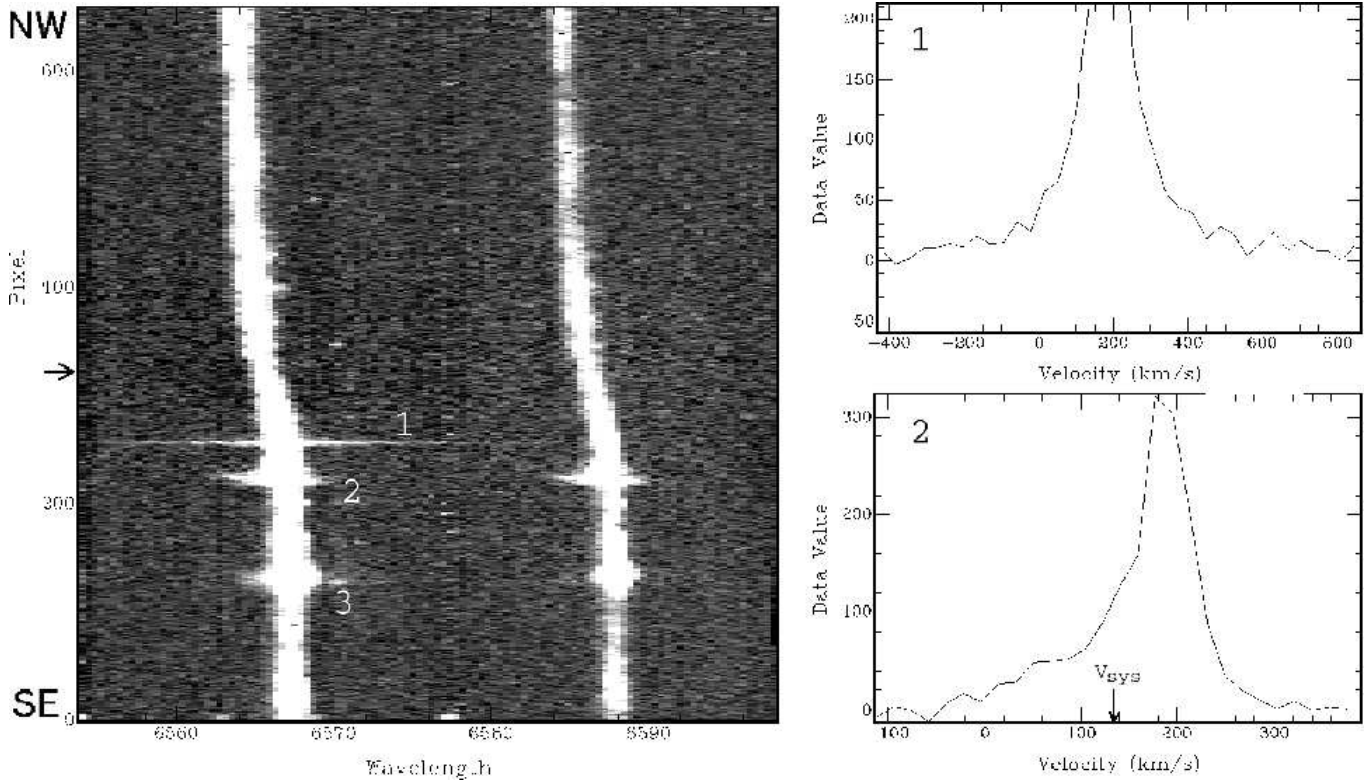}
\caption{
Left panel:
Optical spectrum along the major axis of NGC\,2403 (slit 5 in 
Figure~\ref{f_slits}) in the spectral range of \ha\ and 
\nii.
Note the large velocity dispersion of features visible in the receding (S-E)
side of the galaxy (bottom of this image).
Right panels:
\ha\ line profiles through the broad features 1 and 2 observed
along the major axis of NGC\,2403 and shown in the left panel.
The velocities are heliocentric.
\label{f_maxc1_z1}}
\end{figure}

The extremely broad \ha\ emission feature shown in
Figure~\ref{f_maxc1_z1} (labeled 1)
was detected in two exposures (observation 6 in Table 1) in the third night of 
the observations.
Other two exposures along the major axis (observation 7), 
which were obtained in the same night with a
small shift of less than 2$''$ North-East, did not show any broadening
in that position while the other features were still present and had
similar shapes and locations. 
Clearly, in contrast with the others, feature 1
is unresolved (size smaller than 10$-$15 pc) and, therefore, probably stellar.
It is located at about 300 pc from the centre of the galaxy and has a
total velocity extent of about 1200 km~s$^{-1}$ (see the line profile
in the upper right panel of Figure~\ref{f_maxc1_z1}).
Its total \ha\ flux (only the wings) is about 1.2 $\times$ 10$^{-15}$ 
erg s$^{-1}$ cm$^{-2}$ (luminosity of 1.4 $\times$ 10$^{36}$ erg s$^{-1}$).
It is to far to be connected with the nucleus of NGC\,2403;
moreover NGC\,2403 does not show any evidence of nuclear activity.
No $Chandra$ X-ray source is found at the position of this feature
(\cite{fra02b}; \cite{sch03}).

The most interesting characteristic of this broad feature is
that the emission is clearly visible in \ha\ but not 
in any other line, except for a
weak signal in H$\beta$ in the blue arm of the ISIS spectrograph.
Velocities of a few thousands \kms\ are indeed observed for ejected
materials from Wolf Rayet stars (\cite{cro00}) or Blue Luminous
Variables (\cite{sta01}).
However, in such stars also Helium (and \nii) lines,
absent here, are usually detected.
Another possibility is that of emission from a supernova envelop
(\cite{chu88}). 
The velocity of about 500$-$1000 km~s$^{-1}$ is typical for the
Sedov phase in the expansion of a supernova envelop and SN
age less than 10$^4$ yrs.
A supernova rate of 0.01 yr$^{-1}$ for NGC\,2403
(\cite{mat97}) would mean about 100 detectable SNs in the entire disk,
consistent with our result, considering that 
our observations sample only about 1 \% percent of the disk.
However, also in the case of a supernova the absence of other emission
lines (e.g.\ \sii) could present a problem.

\section{Conclusions}

We have discussed optical long-slit spectroscopy of the 
spiral galaxy NGC\,2403.
The purpose of these observations was to study the kinematics of
the ionised gas and to compare it with the \hi\ results by
Fraternali et al.\ (2001, 2002a) which show the presence of an
extended \hi\ halo rotating more slowly (lagging) with respect to the
disk. 

The results from this optical study of NGC\,2403 can be summarized as follows:

\begin{enumerate}
\item{
The overall kinematics of neutral and ionised gas as inferred from the
\hi\ line and optical recombination lines (\ha\ and \nii) agree
remarkably well.
The \hi\ and \ha\ rotation curves agree also in the very inner
regions of the galaxy.
}
\item{
The slit spectra along the major axis show tails of emission
at velocities lower than rotation (towards systemic).
Such tails are produced by a large-scale lagging component of ionised 
gas which appears to be the ionised counterpart of the lagging \hi\
halo.
This interpretation is confirmed by the presence of a gradient in the
\nii/\ha\ intensity ratio with decreasing rotation velocity.
The lagging halo gas has already been observed in edge-on galaxies
(DIG) but this is the first detection in a galaxy viewed at
intermediate inclination. 
}
\item{
Slits along the minor axis of the galaxy
seem to indicate a large-scale radial inflow of the ionised gas
towards the centre of the galaxy (as found for \hi). 
If the present results are correct, this would imply that the DIG does 
not represent the outflowing component of the fountain.
However, because of the faintness of the ionised tails 
and the poor velocity resolution, a firm conclusion is not possible.
}
\item{
On small scales, there is 
clear evidence for powerful vertical outflows of ionised gas.
Such outflows are detected as local broadenings of the line profiles of up to
300 km~s$^{-1}$.
Such broad features are probably produced by large vertical motions and
originate in regions of ongoing star formation like compact \hii\ regions.
In our observations we have detected two such features, but we have sampled
only a tiny fraction ($\leq$ 1\%) of the galactic disk.
In the hypothesis of a galactic fountain, in order
to match the observed \hi\ infall rate one would need
tens of thousands of such outflows over the all disk.
}

\item{
Also, there are implications for the study of our Galaxy. 
As already pointed out by Fraternali et al.\ (2001, 2002a),
the halo \hi\ features found in external galaxies are probably the
analogue of some of the galactic HVCs. 
Recently, ionised counterparts of several HVCs and also ionised HVCs
with no neutral gas have been found (\cite{ric03}).
The present detection of an ionised counterpart of the \hi\
halo in NGC\,2403 showing a kinematics very similar to that of the
\hi, strengthens this analogy with the galactic HVCs.
}
\end{enumerate}

\begin{acknowledgements}

We are grateful to an anonymous referee for various important improvements to the paper. 

\end{acknowledgements}

\newpage

\clearpage

\begin{table}[ht]
\begin{center}
\begin{tabular}{ccccccc}

\hline
\noalign{\smallskip}

N. & Slit & Location & Date & Number of & Exp.\ Time & r.m.s.\ noise\\
   &  & &  & exposures & (hrs:min) & (erg s$^{-1}$cm$^{-2}$\AA$^{-1}$pix$^{-1}$)\\

\noalign{\smallskip}
\hline
\noalign{\smallskip}

1 & 1   & major N-W     & 2-Jan-01  & 4 & 3:00  & 7.96~$\times$~10$^{-19}$      \\
2 & 2   & minor N-E     & 3-Jan-01  & 2 & 1:30  & 1.22~$\times$~10$^{-18}$      \\
3 & 3   & major S-E     & 3-Jan-01  & 5 & 3:45  & 7.89~$\times$~10$^{-19}$      \\
4 & 4   & minor S-W     & 2-Jan-01  & 3 & 2:15  & 8.28~$\times$~10$^{-19}$      \\
5 &     & minor S-W     & 3-Jan-01  & 2 & 1:30  & 1.35~$\times$~10$^{-18}$      \\
6 & 5   & major centred & 4-Jan-01  & 2 & 1:30  & 1.67~$\times$~10$^{-18}$      \\
7 &     & major centred & 4-Jan-01  & 2 & 1:30  & 1.42~$\times$~10$^{-18}$      \\
8 & 6   & minor centred & 4-Jan-01  & 4 & 3:00  & 1.19~$\times$~10$^{-18}$      \\
\noalign{\smallskip}
\hline
\end{tabular}
\caption[]{Parameters for the spectroscopic observations of NGC2403.}
\label{t_optobs}

\end{center}
\end{table}

\end{document}